\documentclass[11pt,a4paper]{article}
\usepackage{epsfig}
\usepackage{authblk}
\usepackage[nohead, nomarginpar, margin=1in, foot=.25in]{geometry}
\usepackage[T1]{fontenc}
\usepackage[utf8]{inputenc}
\usepackage{latexsym}
\usepackage{subfig}
\date{}
\newcommand{\sn}{\sqrt{s_{\rm{NN}}}}
\newcommand{\s}{\sqrt{s}}
\newcommand{\pt}{\textit{p}_{\rm{T}}}

\newcommand{\dnchdeta}{\langle \rm d \textit{N}_{ch} / d {\eta} \rangle} 
\begin{document}
\makeatletter
  \def\title@font{\Large\bfseries}
  \let\ltx@maketitle\@maketitle
  \def\@maketitle{\bgroup%
    \let\ltx@title\@title%
    \def\@title{\resizebox{\textwidth}{!}{%
      \mbox{\title@font\ltx@title}%
    }}%
    \ltx@maketitle%
  \egroup}
\makeatother
\title {\bf Study of strange non-strange hadron ratios in pp and p-Pb collisions at LHC energies}
\author[1,2,\footnote{ssarita.2006@gmail.com}]{S. Sahoo}
\author[$$]{R. C. Baral}
\author[1,2]{P. K. Sahu}
\author[2]{M. K. Parida}
%\author[1,2] {et. al.}
%
%
\affil[1] {Institute of Physics, HBNI, Bhubaneswar, India}
\affil[2] {CETMS, Siksha `O' Anusandhan Deemed to be University, Bhubaneswar, India}
\maketitle
\begin{abstract}
It has been observed that the yields of strange and multi-strange hadrons relative to pion increase significantly with the event charged-particle 
multiplicity. We notice from experimental data that yield ratios between non-strange hadrons, like p/$\pi$ or hadrons of same strange content, like $\Lambda$/K$_s^0$, show similar 
enhancement. We have studied this behavior within the ambit of a parton model (EPOS3) and A Multi-Phase Transport (AMPT) model in pp and p-Pb collisions 
at LHC energies. We investigate model predictions of yields and yield ratios of different identified hadron productions as a function of charged-particle 
multiplicity and compare them with published ALICE results. The string melting versions of AMPT and EPOS are found to establish enhancements in the particle 
yield ratios.

\iffalse
We compare yields and the ratios of different identified particles produced in the models with ALICE data. 
Though models deviate from data, the string melting version of AMPT and EPOS establish the enhancement in the particle ratios.  \fi
%
\end{abstract}
\section{Introduction}
The production of strange hadrons in the high energy hadronic interactions provides the effective tools to investigate the properties of Quark-Gluon Plasma (QGP).
The strangeness is produced in hard partonic scattering processes by flavor creation, flavor excitation and gluon splitting. These
productions dominate in high $\pt$ regions. The non-perturbative processes, like string fragmentation, dominate the productions
at low $\pt$ regions. The relative yields of strange particles to $\pi$ in heavy ion collisions from top RHIC to LHC energies are found to be
compatible with those of a hadron gas in thermal and chemical equilibria, and this behavior can be described using grand-canonical statistical
models \cite{Statistical, Statistical2}. In peripheral collisions, the relative yields of strange particles to $\pi$ decrease and tend toward
those observed in pp collisions. This behavior can be described by statistical mechanics approach. The statistical models, by implementing
strangeness canonical suppression, can predict a suppression of strangeness production in small systems. It has been seen in ALICE for pp collisions 
at $\sqrt{s}$ = 7 TeV that the $\pt$ integrated yields of strange and multi-strange particles relative to $\pi$ increase significantly with
charged-particle multiplicity \cite{Nature}. The observed relative yield enhancements increase with the strange quark content in hadrons. Further, it will be 
interesting to study the behavior of ratios between same strange content particles. This study may open up a new possibility of particle production mechanism 
in hadron collisions at microscopic level.
\paragraph{}
In this paper, we present the comparison of yields and the respective normalized ratios of identified particles between ALICE data and the models like AMPT (A Multi-Phase 
Transport Model) \cite{Z.Lin} and EPOS (3+1 Hydrodynamics Model) \cite{KWerner, KWerner2}. In section 2, we have described detail of the models. We discuss
the results from models along with ALICE data in section 3 and the summary is given in section 4.
%the yield enhancements behavior with charged-particle multiplicity using models like AMPT (A Multi-Phase Transport Model) \cite{Z.Lin} and EPOS (3+1 Hydrodynamics Model) \cite{KWerner}.
%
\section{Description of models}
%\subsection{The AMPT Model}
The AMPT model (version 1.26t7/2.26t7) mainly provides the initial conditions, partonic interactions, conversion from the partonic to hadronic matter, and
hadronic interactions. The initial conditions, which include the spatial and momentum distributions of minijet partons and 
string excitations, are obtained from the HIJING Model \cite{HIJ1,HIJ2,HIJ3,HIJ4}. The partonic interactions or the scatterings
among partons are modeled by Zhang's parton cascade (ZPC) \cite{Zhang}. Here, we have studied both the versions of AMPT model, i.e.,
default version and string-melting (SM) version. In the default AMPT model, partons are recombined with their parent strings 
when they stop interacting, and the resulting strings are converted to hadrons using the Lund string fragmentation model \cite{Lund1,Lund2}.
In the AMPT SM, a quark coalescence model is used instead to combine partons into hadrons. The dynamics of
the subsequent hadronic matter is described by a relativistic transport (ART) model \cite{ART1,ART2} and extended to
include additional reaction channels which are important in high energy processes, like formation and decay of resonances, baryon and antibaryon
production from mesons and their inverse reactions of annihilation. Final results from the AMPT model are obtained after hadronic interactions are
terminated at a cut off time ($t{_{cut}}= 40fm/c$) when observables under study are considered to be stable, i.e., when further hadronic interactions 
after $t{_{cut}}$ will not significantly affect these observables. As per model recommendation, default set of parameters are used for this study. 

The model EPOS (version 3.107) is a parton model, with many binary parton-parton interactions, each one creating a parton ladder. EPOS describes the full 
evolution of a heavy-ion collisions. The initial stage is a multiple scattering approach based on pomerons and strings. The reaction volume is composed of 
two parts, i.e., core and corona. The core provides the initial condition for the QGP evolution, hence one employs viscous hydrodynamics.
Hadrons from string decays simply form the corona part. After hadronization the core and corona hadrons are fed into UrQMD \cite{MicroscopicModel}, which describe hadronic interactions in microscopic approach. The chemical and 
kinetic freeze-outs occur within this phase. EPOS is designed to be used for particle physics experimental analysis (SPS, RHIC, LHC) for pp or heavy ion from 
100 GeV to 1000 TeV energies. Details of EPOS model can be found in Ref. \cite{KWerner, KWerner2} and references therein. 

In the ALICE experiment, multiplicity and centrality determination are done via the measurement of the charged-particle multiplicity using two forward detectors 
V0A and V0C \cite{ALICE_V0}, which span the pseudorapidity ranges 2.8 $< \eta <$ 5.1 and -3.7 $ < \eta < $-1.7 in the ALICE frame of experiment, respectively. 
Charged-particle multiplicities in the pseudorapidity ranges of both the detectors are used to classify multiplicity classes in pp collisions. In p-Pb collisions, 
centrality selection is done via the measurement of charged-particle multiplicity in V0A acceptance, which is Pb going direction. The mean charged-particle 
multiplicity densities ($\dnchdeta$), corresponding to each multiplicity or centrality classes, are calculated in the pseudorapidity 
ranges $|\eta| < 0.5 $. The above procedure is followed to adopt the steps pursued by ALICE \cite{ALICE_Cent}.
\begin{figure}[!h]
\vspace{-0.4cm}
\centering
\includegraphics[scale=0.45]{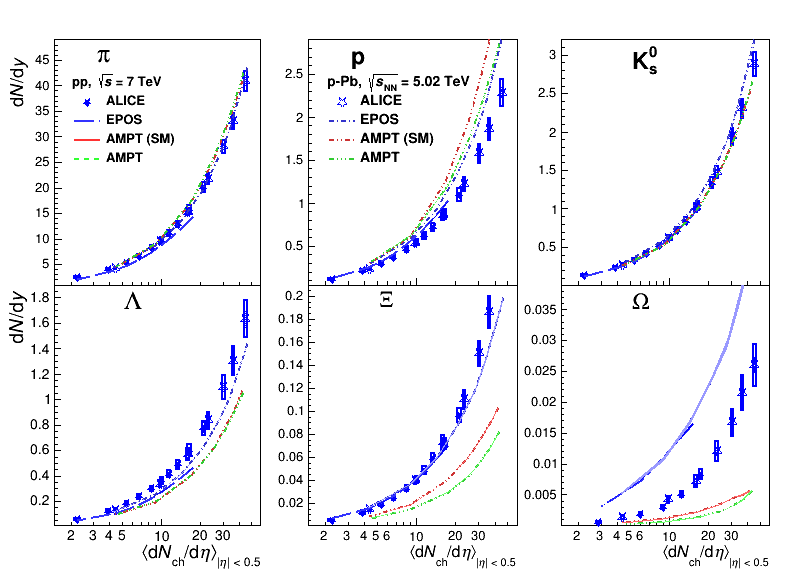}
\caption{Yields of hadrons vs charged-particle multiplicity measured in pp collisions at $\sqrt{s}$ = 7 TeV and p-Pb collisions at $\sn$= 5.02 TeV. The markers show ALICE data \cite{Nature} and the lines show results from models and their statistical uncertainties are shown by shaded bands. }
\label{fig:1}
\end{figure}
\begin{figure}[!b]
\vspace{-0.4cm}
\centering
\includegraphics[scale=0.5]{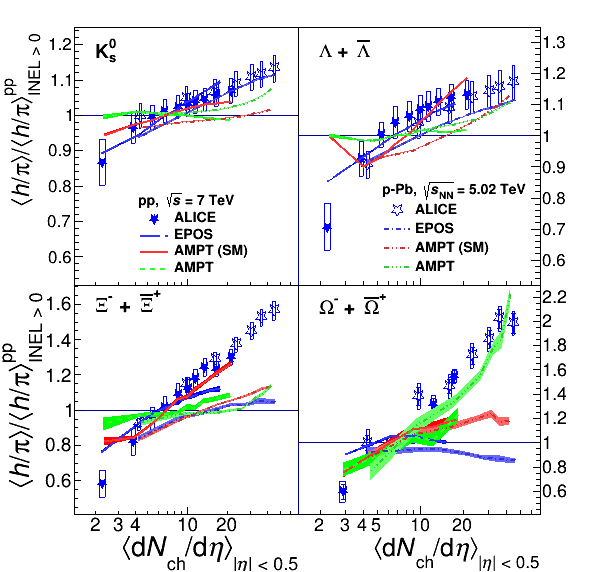}
\caption{Yield ratios of hadrons to $\pi$ in pp collisions at $\sqrt{s}$ = 7 TeV and p-Pb collisions at $\sn$= 5.02 TeV, normalized to the values measured in the inclusive INEL $>$ 0 pp collisions at $\sqrt{s}$ = 7 TeV. The solid and open star symbols show ALICE data \cite{Nature} in pp and p-Pb collisions, respectively.  The lines show results from models and shaded bands show their statistical uncertainties.}
\label{fig:2}
\end{figure}
%
%
%  
%##################################################
%##################################################
\section{Results and Discussion}
%%%%%%%%%%%%%%%%%%%%%%%%%%%%%%%%
We have simulated AMPT and EPOS events in pp collisions at $\s$ = 7 TeV and in p-Pb collisions at $\sn$ = 5.02 TeV.
The Fig.\ref{fig:1} shows the primary yields of hadrons, like $\pi$, p, $\rm{K_{s}^{0}}$, $\rm{\Lambda}$, $\rm{\Xi}$, and $\rm{\Omega}$ as a function of 
event charged-particle multiplicity density ($\dnchdeta$). For bravety, $\pi^+ + \pi^-$, p + $\overline{p}$, $\Lambda + \overline{\Lambda}$, 
$\Xi^- + \overline{\Xi}^{_+}$ and $\Omega^- + \overline{\Omega}^{_+}$ are denoted as $\pi$, p, $\Lambda$, $\Xi$, and $\Omega$, respectively. Here a primary particle is defined as a particle created in the collision, but 
not coming from a weak decay. The experimental data points are shown by markers in the figure and different model results are shown by respective colored lines. 
The statistical uncertainities from model results are shown by shaded bands. We note that the EPOS model describes quantitatively most of experimental data points within 
two standard deviations, but it highly overestimates the $\rm{\Omega}$ yields. We further find that the AMPT default and AMPT SM models well describe the 
experimental data points for mesons, but underestimate yields for strange baryons. The AMPT SM gives higher yield for baryons when compared to the AMPT default. The 
higher is the strange baryon mass, the lower is the AMPT model prediction from experimental data as a function of charged-particle multiplicity density.
The Fig.\ref{fig:2} shows primary yield ratios of hadrons ($\rm{K_{s}^{0}}$, $\rm{\Lambda}$, $\rm{\Xi}$, and $\rm{\Omega}$) to $\pi$ yield 
(denoted as $\langle h/\pi \rangle $) divided by the same values measured in the inclusive inelastic pp collisions ($\langle h/\pi \rangle^{\rm pp}_{\rm INEL > 0} $ ) as a function of $\dnchdeta$, calculated from EPOS and 
AMPT models. The figure also shows same ratios from ALICE data. Here INEL $>$ 0 represents events having at least one charged 
particle produced in the pseudorapidity interval $|\eta| < 1.0 $. The measurements were performed at mid-rapidity $|y| < 0.5$ in pp collisions and within
$-0.5 < y <0$ in p-Pb collisions at the center of mass frame. 
A detailed information of the ALICE data can be found in Ref. \cite{Nature} and references therein. From Fig.\ref{fig:2}, it is clearly seen that EPOS qualitatively 
explains enhancements in the $\langle h/\pi \rangle $ ratios as a function of $\dnchdeta$ with an exception for $\rm{\Omega}$. The EPOS gives a higher rate of 
production for $\pi$ as a function of $\dnchdeta$ compared to $\rm{\Omega}$. AMPT SM describes the enhancement in the $\langle h/\pi \rangle $ ratios 
as a function of $\dnchdeta$. The enhancement in the $\rm{\Omega}/\pi$ ratios in p-Pb collisions is noted to be well described by the AMPT model where Lund string fragmentation
is incorporated for hadronization. 
\begin{figure}[!h]
\vspace{-0.4cm}
\centering
\includegraphics[scale=0.5]{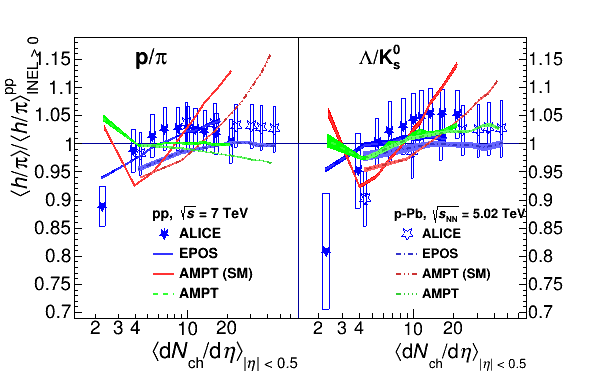}
\caption{Yield ratios p/$\pi$ and $\rm \Lambda/K_{s}^{0}$  in pp collisions at $\sqrt{s}$ = 7 TeV and p-Pb collisions at $\sn$= 5.02 TeV, normalized to the values measured in the inclusive INEL $>$ 0 pp collisions at $\sqrt{s}$ = 7 TeV. The solid and open star symbols show ALICE data \cite{Nature} in pp and p-Pb collisions, respectively.  The lines show results from models and shaded bands show their statistical uncertainties.}
\label{fig:3}
\end{figure}

In Fig.\ref{fig:3} we have displayed the above said normalized ratios between non-strange hadrons, $\rm p/\pi$ and single-strange hadrons, 
$\rm \Lambda/K_{s}^{0}$ as a function of charged-particle multiplicity density. Here we notice from experimental data that the ratios between hadrons of same strange content show similar enhancement as a function $\dnchdeta$
like previously shown  $\langle h/\pi \rangle $ ratios in Fig.\ref{fig:2}. The AMPT SM and EPOS clearly depict a sign of enhancement in these hadron ratios
as a function of $\dnchdeta$. In AMPT SM version, the model need to melt the strings into partons for energy density beyond a certain critical value. It may be 
difficult to achieve that critical energy density when one approaches to very low $\dnchdeta$. This could be a possible reason for AMPT SM to approach towards
AMPT default version at low $\dnchdeta$. This behaviour can be seen for ratios like $\rm \Lambda/\pi$, $\rm p/\pi$, and $\rm \Lambda/K_{s}^{0}$.
\section{Summary}
We have presented yields of strange and non-strange hadrons at mid-rapidity and their ratios in AMPT and EPOS models in pp collisions at $\s$ = 7 TeV and in 
p-Pb collisions at $\sn$ =  5.02 TeV. We have compared model results with ALICE data. We observed that, like experimental data, models qualitatively described the 
enhancement in the yields of hadrons at mid-rapidity with event charged-particle multiplicity density. Both the versions of AMPT model are found to underestimate yields of 
strange baryons, whereas EPOS highly overestimates $\Omega$ yield. The AMPT SM and EPOS show multiplicity-dependent enhancement of the production of strange 
hadrons relative to $\pi$. The Lund string fragmentation implemented version of AMPT is found to be quite successful in describing the multiplicity-dependent 
enhancement in $\Omega/\pi$ ratios. The models predict multiplicity-dependent enhancement in the yield ratios between a heavier baryon and a lighter meson of 
same strange content, though there are quantitatively deviation from experimental data. This behaviour is completely independent of strangeness of hadrons. One needs more 
experimental data with improved uncertainties to figure out whether this is a mass-dependent phenomenon. 

\end{document}